\documentclass[twocolumn,aps,epsfig,nofootinbib]{revtex4}

\usepackage{graphicx}
\usepackage{amsfonts}
\usepackage{epstopdf}
\usepackage{latexsym}
\usepackage{amssymb}
\usepackage{amssymb}

\usepackage[center]{subfigure}

\begin{document}

 \newcommand{\bq}{\begin{equation}}
 \newcommand{\eq}{\end{equation}}
 \newcommand{\bqn}{\begin{eqnarray}}
 \newcommand{\eqn}{\end{eqnarray}}
 \newcommand{\nb}{\nonumber}
 \newcommand{\lb}{\label}
\newcommand{\PRL}{Phys. Rev. Lett.}
\newcommand{\PL}{Phys. Lett.}
\newcommand{\PR}{Phys. Rev.}
\newcommand{\CQG}{Class. Quantum Grav.}

\title{
Stationary axisymmetric and  slowly rotating spacetimes in  Ho\v{r}ava-Lifshitz  gravity}

 \author{Anzhong Wang}
\email{anzhong_wang@baylor.edu}

\affiliation{Institute  for Advanced Physics $\&$ Mathematics,   Zhejiang University of
Technology, Hangzhou 310032,  China\\
GCAP-CASPER, Physics Department, Baylor
University, Waco, TX 76798-7316, USA }

\date{\today}

\begin{abstract}
 
Stationary, axisymmetric  and  slowly rotating vacuum spacetimes in  the Ho\v{r}ava-Lifshitz (HL) gravity are studied, 
and shown that, for any given spherical static vacuum solution of the HL theory (of any model, including the ones with 
an additional U(1) symmetry), there always exists a corresponding slowly rotating, stationary and axisymmetric vacuum 
solution, which reduces to the former,  when the rotation is switched off.  The rotation is universal and only implicitly 
depends on the models of the HL theory and their coupling constants  through the spherical seed solution. As a result, 
all asymptotically flat slowly rotating vacuum solutions are asymptotically identical to the slowly rotating Kerr solution. 
This is in contrast to the claim of Barausse and Sotiriou, Phys. Rev. Lett. {\bf 109}, 181101 (2012), in which slowly rotating 
black holes were reported (incorrectly) not to exist in the infrared limit of the non-projectable HL theory.

\end{abstract}


\maketitle

{\em Introduction}:---Since Einstein proposed his general relativity (GR) in 1915,  various  experiments and observations have been 
carried out, and  so far  all of them are consistent with it  \cite{Will05}. Despite of these splendid achievements, it has been realized for 
a long time that GR is not (perturbatively) renormalizable \cite{HV}, and thus may not be used to describe quantum effects of
gravity in very short distances. On the other hand, because of the universal coupling of gravity to all forms of energy, it is expected that
 it should have a quantum mechanical description. Motivated by this anticipation, quantization of gravitational fields has been one of the
main driving forces in physics in the past decades \cite{QGs}.

Recently, Ho\v{r}ava \cite{Horava} proposed a theory of quantum gravity within the framework of quantum field theory, in which the 
fundamental variables are the metric. One of the essential ingredients of the theory is the inclusion of higher-dimensional spatial
derivative operators, so that the ultraviolet (UV) behavior is dominated by them and that the  theory is power-counting
renormalizable. The exclusion of higher-dimensional time derivative operators, on the other hand,  guarantees that the theory is unitary,
a problem that has been faced  in quantization of gravity for a long time \cite{Stelle}. However, this inevitably breaks  Lorentz symmetry in 
the UV. Although  such a breaking  in the gravitational sector is much less restricted by experiments/observations than that
in the matter sector  \cite{LZbreaking} (See also \cite{Pola}), it is still a challenging question how to prevent the propagation of the 
Lorentz violations into the Standard Model of particle physics \cite{PS}.  In the infrared (IR) the lower dimensional operators
take over, presumably providing a healthy low energy limit.

The Lorentz breaking in the UV   are realized by invoking the anisotropic scaling between time and space, 
$t \rightarrow b^{-z} t,\; \vec{x} \rightarrow b^{-1}\vec{x}$. This is a reminiscent of Lifshitz scalars \cite{Lifshitz} in condensed matter physics, 
hence the theory is often referred to as the HL gravity.  To be power-counting renormalizable, the critical exponent $z$ has to be $z \ge 3$
\cite{Horava,Visser}.  Ho\v{r}ava assumed that the symmetry is broken only down to the so-called foliation-preserving diffeomorphism,
\begin{equation}
\lb{0.2}
 t' = f(t), \;\;\; {x'}^i = \zeta^i(t, x),  
\end{equation}
 denoted often by Diff($M, \; {\cal{F}}$). With such a breaking,   spin-0 gravitons in general appear, in addition to the spin-2 ones found in GR.
This is potentially dangerous, and leads to several problems, such as instability,  strong coupling and different speeds of massless particles 
\cite{reviews}. To resolve these problems, various models have been proposed, including the healthy extension of the non-projectable 
HL theory \cite{BPS}, and a more dramatical modification, in which  an extra local U(1) symmetry is introduced, so that the symmetry
of the theory is enlarged to  \cite{HMT},
\bq
\lb{symmetry}
 U(1) \ltimes {\mbox{Diff}}(M, \; {\cal{F}}).
\eq
Because of this extra symmetry,  the spin-0 gravitons are eliminated \cite{HMT,WWa}. As a result, all the problems related to them, including the ones
mentioned above,  are resolved. This was initially done with the projectability condition \cite{HMT,WWa}, in which     the lapse function 
 in the Arnowitt-Deser-Misner  (ADM) decompositions \cite{ADM} is a function of $t$ only. Soon, it was  extended to the case without it \cite{ZWWS}. 

In this Letter, we shall investigate  another important issue of the HL theory: Stationary axisymmetric and slowly rotating gravitational fields of 
black holes and stars.  The existence of such fields are fundamental to the theory, since rotating objects are more common than non-rotating ones 
in our universe.  In particular, observations show that rotating black holes  very likely exist \cite{Nara}. Certainly, the issue of black holes in the HL 
theory is very subtle, because of the Lorentz violations and modified dispersion relationship,
\bq
\lb{0.3}
E^2 = c_{p}^2 p^2\left(1 + \alpha_1 \left(\frac{p}{M_{*}}\right)^2 +  \alpha_2  \left(\frac{p}{M_{*}}\right)^4\right),
\eq
where $E$ and $p$ are the energy and momentum of the particle considered, and $c_p, \; \alpha_i$ are coefficients, depending on the particular 
specie of the particle, while $M_{*}$ denotes the suppression energy  scale of the higher-dimensional operators. Then,   one can see that 
both phase and group velocities of the particles are unbounded  with the increase of energy. This suggests that black holes may not exist at all in the 
HL theory. However, in the IR the high-order terms of $p$ are negligible, and the first term in Eq.(\ref{0.3}) becomes dominant, so one may still define 
black holes, following what was done  in GR. For more detail, we refer readers to \cite{GLLSW} and references therein. 

With the above in mind, in this Letter we shall show that for any given spherical static vacuum solution of the HL theory, there always exists a corresponding  
stationary  axisymmetric and slowly rotating vacuum solution, which  reduces to the former,  when the rotation is switched off. This is true in all the models 
of the HL theory proposed so far, including the ones with the enlarged symmetry (\ref{symmetry}).  In addition, the rotation is universal and only implicitly 
depends on the   models of the HL theory and their coupling constants  through the spherical seed solution [cf. Eq.(\ref{2.6})]. As a result, all asymptotically 
flat slowly rotating vacuum solutions are asymptotically identical to the slowly rotating Kerr solution found in GR, given by Eq.(\ref{2.7}).

{\em HL Theory without U(1) Symmetry}:---A naturally starting point to formulate the HL theory is the ADM  decompositions \cite{ADM}, 
$\left(N, N^i, g_{jk}\right)$, where $N, \; N^i$ and $g_{jk}$ are, respectively, the lapse function, shift vector, and the 3-dimensional  metric
defined on the leaves   $t = $ Constant.  Then, the general  action  takes the form \cite{BPS},
\bq
\lb{1.2}
S_{g} =\frac{1}{16\pi G} \int{dt d^3x N\sqrt{g}\left({\cal{L}}_{K} - {\cal{L}}_{V}\right)},\nb
\eq
 where   $ {\cal{L}}_{K} \equiv K_{ij}K^{ij} - \lambda K^2$, and
${\cal{L}}_{V} = {\cal{L}}_{V}\big(a_i, R_{jk},\; \nabla_{l}\big)$ denotes the potential made of all the operators constructed from 
$a_{i}, R_{ij}$ and $\nabla_i$. Its explicit form is irrelevant to the current problem,  so we shall not present it here.  $R_{ij}$ 
denotes the Ricci tensor of $g_{ij}$,   $\nabla_{i}$  the covariant derivative with respect to $g_{ij}$, and $ a_i \equiv {N_{,i}}/{N}$.  
$K_{ij}$ is the extrinsic curveture of the leaves $t = $ Constant, given by  $ K_{ij} = \left(- \dot{g}_{ij} + \nabla_i N_j +  \nabla_j N_i\right)/(2N)$.  
In addition, in the case with the projectability  condition, we have $N = N(t)$, and then $a_{i} = 0$. In this Letter, we shall treat the case with the 
projectability condition as a particular case of the one without it, $N = N(t, x)$, whenever it is possible.  

Then, variation of $S_g$ with respect to $N$ yields the Hamiltonian constraint,  
\bq
\lb{equ1}
(i) \;\; \int{d^3x \sqrt{g} {\cal{H}}^{\perp}} = 0,  \;\;\;\;\;
(ii) \;\; {\cal{H}}^{\perp} = 0, 
\eq
respectively, for the projectable and non-projectble cases,  where ${\cal{H}}^{\perp} \equiv {\cal{L}}_{K} + {\cal{H}}$,
${\cal{H}} \equiv {\delta\left(N{\cal{L}}_{V}\right)}/{\delta N} = {\cal{H}}\big(a_i, R_{jk},\; \nabla_{l}\big)$. Variation of $S_g$ with respect to $N^i$ 
yields the momentum constraint,
\bq
\lb{equ2}
\nabla^{j}\pi_{ij} = 0,
\eq
where $\pi_{ij} \equiv - K_{ij} + \lambda g_{ij} K$. 
Finally, the variation of $S_g$ with respect to $g_{ij}$ yields the dynamical equations, 
\bqn
\lb{equ3}
&& \frac{1}{N\sqrt{g}} \frac{\partial}{\partial t}\left(\sqrt{g} \pi^{ij}\right) = \frac{1}{N}\left[\nabla_k\left(\pi^{ij}N^k\right) - 2\pi^{k(i}\nabla_k N^{j)} \right]   \nb\\
&& ~~~~~ - 2\left(K^{ik}K^j_k-\lambda K K^{ij}\right) + F^{ij} + \frac{1}{2}g^{ij}{\cal{L}}_K, ~~~
\eqn
where $ F^{ij} \equiv  - \left[{\delta \left(N\sqrt{g}{\cal{L}}_{V}\right)}/{ \delta g_{ij}}\right]/({N\sqrt{g}}) = F^{ij}\big(a_{k}, R_{lm}, \nabla_{n}\big)$.

{\em Slowly Rotating Spacetimes}:---Let us consider the gravitational field of a body with slow and uniform rotation about an axis described by, 
\bqn
\lb{SRM}
N &=&  {N}(r), \;\; N^{i} = {h}(r) \delta^{i}_{r}  +  \omega(r, \theta) \delta^{i}_{\phi}, \nb\\
  g_{ij} &=&   {\mbox{diag.}}\left(f^{-1}(r), r^2, r^2\sin\theta^2\right),
\eqn
where slow rotation means $|\omega| \ll 1$.   Thus, one can consider the gravitational field  of a slowly rotating body as linear perturbations over the
spherical background, $\left(N(r),  f(r), h(r)\right)$. Then, to the first-order of $\omega$, we obtain
\bqn
\lb{2.1a}
 a_{i}  =       \bar{a}_{i},\; R_{ij} = \bar{R}_{ij},\; \nabla_{i} = \bar{\nabla}_{i},\;
K_{ij}  = \bar{K}_{ij} + \delta K_{ij},
\eqn
where  quantities with bars denote the ones calculated from the spherical seed solutions $(N, f, h)$, and
\bqn
\lb{2.1}
\bar{K}_{ij} &=& \frac{1}{2{N}{f}^2}\left(2{f}{h}' - {h}{f}'\right)\delta^{r}_{i} \delta^{r}_{j} + \frac{r{h}}{{N}}\Omega_{ij},\nb\\
\delta K_{ij} &=&  \frac{r^2\sin^2\theta}{2{N}}\Big(\omega_{,i}\delta^{\phi}_{j} + \omega_{,j}\delta^{\phi}_{i}\Big),
\eqn
with $\Omega_{ij} = \delta^{\theta}_{i} \delta^{\theta}_{j} + \sin^2\theta  \delta^{\phi}_{i} \delta^{\phi}_{j}$, and $f' \equiv \partial f/\partial r$, etc.  
It is important to note that $ a_i, \; R_{ij}$ and $\nabla_{i}$ all do not contain terms of first-order of $\omega$. This is simply because that
$N$ and $g_{ij}$ do not contain such terms. As a result, the potential ${\cal{L}}_{V}\big(a_i, R_{jk},\; \nabla_{l}\big)$ and terms resulted from it, such as 
${\cal{H}}\big(a_i, R_{jk},\; \nabla_{l}\big)$ and $F^{ij}\big(a_i, R_{jk},\; \nabla_{l}\big)$,  also do not contain such terms, that is
${\cal{H}} = \bar{\cal{H}} + {\cal{O}}(\omega^2),\; F^{ij} = \bar{F}^{ij} + {\cal{O}}(\omega^2)$. Moreover, since 
\bq
\lb{2.2} 
\delta K = \bar{g}^{ij}\delta{K}_{ij} = 0,\;\;\; \bar{K}^{ij}\delta{K}_{ij} = 0, 
\eq
we find that ${\cal{L}}_{K}$ does not contain terms of first-order of $\omega$
either. Then,  ${\cal{L}}_{V}  = \bar{\cal{L}}_{V}\big(\bar{a}_i, \bar{R}_{jk},\; \bar{\nabla}_{l}\big)  + {\cal{O}}(\omega^2)$, and
 ${\cal{L}}_{K}   = \bar{\cal{L}}_{K} + \delta{\cal{L}}_{K}$, where
\bqn
\lb{2.3}
\bar{\cal{L}}_{K} &=& \frac{1}{8r^2{N}^2{f}^2}\Big[(1-\lambda)r^2\big({h}{f}' - 2{f}{h}'\big)^2 \nb\\
&& 
+ 8 (1-2\lambda){f}^2{h}^2 + 8\lambda r {f}{h}\big({h}{f}' - 2{f}{h}'\big)\Big],\nb\\
 \delta{\cal{L}}_{K} &=& \frac{\sin^2\theta}{2{N}^2}\Big(r^2{f} \omega_{,r}^2 + \omega_{,\theta}^2\Big).
\eqn

To zeroth order of $\omega$, Eqs.(\ref{equ1}), (\ref{equ2})  and (\ref{equ3}) yield the HL field 	equations for  $\left(N(r),  f(r), h(r)\right)$. 

When $h = 0$, from Eq.(\ref{2.1}) - (\ref{2.3}) we can see that  $\bar{K}_{ij}  = \bar{\cal{L}}_{K} = 0$. Thus, to firs-order of $\omega$, Eqs.(\ref{equ1}) 
and (\ref{equ3}) are satisfied identically, simply because there are no non-vanishing terms of the first order of $\omega$ in these equations. On the 
other hand,  we have  $\pi_{ij} =   - \delta{K}_{ij}$ where $\delta{K}_{ij}$ is given by Eq.(\ref{2.1}).  Inserting it into Eq.(\ref{equ2}), we find that it has 
only one independent equation, given by, 
\bq
\lb{equ4}
\frac{\sqrt{{N}^2{f}}}{r^2}\left(r^4 \sqrt{\frac{{f}}{{N}^2}}\; \omega'\right)' + \frac{1}{\sin^3\theta}\left(\sin^3\theta \omega_{,\theta}\right)_{,\theta} = 0.
\eq
It is easy to show that   non-singular solutions for any $\theta$ exist only when $\omega = \omega(r)$. Then, we find that
\bq
\lb{2.6}
\omega(r) = - 3J\int{\sqrt{\frac{N^2}{f}} \;\; \frac{dr}{r^4}} + \omega_0,
\eq
where $J$ and $\omega_0$  are constants. Without loss of the generality, we can  set $\omega_0 = 0$ by the coordinate transformation
$\phi \rightarrow \phi + \omega_0 t$.
 
When $h \not=0$,  we find that to first-order of $\omega$,  the Hamiltonian constraint Eq.(\ref{equ1}) is satisfied identically with the same arguments 
as those given above for the case $h = 0$.  On the other hand, to first-order of $\omega$, the momentum constraint
(\ref{equ2})  yields the same equation (\ref{equ4}). For the dynamical equations (\ref{equ3}), to the
first-order of $\omega$, only the first two terms on the right-hand side of 
Eq.(\ref{equ3}) now are not zero. Taking Eq.(\ref{equ2}) into account, we find that these two terms have two 
non-vanishing components, ($r, \theta$) and ($r, \phi$), and can be cast in the forms,  
\bqn
\lb{equ5a}
&&  \left(r^4 \sqrt{\frac{{f}}{{N}^2}}\; \omega'\right)' = 0,\\
\lb{equ5b}
&&  \left(\frac{r^2 h \omega}{N\sqrt{f}}\right)'_{,\theta} = 0.
\eqn
The combination of Eqs.(\ref{equ4}) and (\ref{equ5a}) immediately yields, $\omega(r, \theta) = \omega_{2}(r)\int{\sin^{-3}\theta  d\theta} 
+ \omega_{1}(r)$,  where $\omega_{2}(r)$ and $\omega_{1}(r)$ are two integration functions of $r$ only. Clearly, $\omega$ becomes singular 
at $\theta = 0, \pi$, unless  $\omega_{2}(r) = 0$. Then, substituting it into Eq.(\ref{equ5a}), we find that $\omega$ is  also given by Eq.(\ref{2.6}). 
Asymptotical flatness condition requires  $f \simeq N^2 \sim 1$ and $h \sim 0$. Then, we find that
\bq
\lb{2.7}
\omega \simeq \frac{J}{r^3},\;\;\; (r \gg 1).
\eq
Therefore, we conclude that, {\em for any given static spherical vacuum seed solution, $\left({N}, {f}, {h}\right)$, of the  HL theory with or without the projectability
condition,  there always exists a slowly rotating vacuum solution $\left({N}, {f}, {h}, \omega\right)$, where $\omega$ is given by   Eq.(\ref{2.6}). When the rotation is
switched off, it reduces to the spherical seed solution}.

Note that our above conclusions hold for the case with any value of $\lambda$, including the one with $\lambda = 1$. In particular,  the slowly rotating Kerr 
solution belongs to it. In fact,  setting  ${\cal{L}}_{V} = - R$ in the general action (\ref{1.2}),   where $R$ denotes the Ricci scalar made of the 3D metric $g_{ij}$, 
 we obtain two solutions, given, respectively, by   $\big({N}^2,  {f},   {h}\big) = (1, 1,  \sqrt{r_g/r}\big)$ and  $\big({N}^2,  {f},   {h}\big) =\big(1 - r_g/r, 1 - r_g/r, 0\big)$, where
 $r_g$ is the Schwarzschild radius.  In  both cases, we have ${f} = {N}^2$. Then, from Eq.(\ref{2.6}), we find that $\omega$ takes the form of Eq.(\ref{2.7})   
 for any $r$. Both of these two seed solutions are  the Schwarzschild solution found in GR,  but written in different coordinate systems 
and are related by  \cite{GPW} $t' = t +  \int{\sqrt{\frac{1-{N}^2}{{f}{N}^2}}\;  dr}$. However,  this kind of coordinate transformations are forbidden by 
the foliation-preserving diffeomorphisms (\ref{0.2}), so in the HL theory these two seed solutions and
their corresponding slowly rotating ones actually represent different spacetimes  \cite{GLLSW}.

 {\em Slowly Rotating Spacetimes with U(1) Symmetry}:---The above conclusion can be easily generalized to the cases with the U(1)
  symmetry \cite{HMT,WWa,ZWWS}. In fact, with the presence of the gauge field $A$ and the Newtonian prepotential
$\varphi$, the action can be still cast in the form (\ref{1.2}), but now with the potential ${\cal{L}}_{V}$ being  replaced by \cite{WWa,ZWWS},
\bqn
\lb{equ11a}
 {\cal{L}}_{V} - {\cal{L}}_{A} - {\cal{L}}_{\lambda,\varphi},\nb
\eqn
where  $ {\cal{L}}_{A} \equiv A(R - 2\Lambda_g)/N$, and $\Lambda_g$ is a coupling constant. 
${\cal{L}}_{\lambda,\varphi} = {\cal{L}}_{\lambda,\varphi} \left(a_i, g_{ij}, R_{ij}, K_{ij}, \varphi, \nabla_i\right)$, where 
  its exact dependence on these variables  is not important to our current discussions, as in the gauge  
\bq
\lb{gauge}
\varphi = 0, 
\eq
to be  chosen below, it vanishes 
identically   \cite{ZWWS,WWa}. 

Then, the Hamiltonian constraint still takes the form of Eq.(\ref{equ1}) but now with ${\cal{H}} \equiv {\delta\left[N({\cal{L}}_{V}- {\cal{L}}_{A} - {\cal{L}}_{\lambda,\varphi})\right]}/{\delta N}
= {\delta\left[N({\cal{L}}_{V}- {\cal{L}}_{A})\right]}/{\delta N} =  {\cal{H}}\big({A}, {a}_i, {R}_{jk}, {\nabla}_{l}\big)$ within the above gauge. Thus, to the first-order of
$\omega$, we have  ${\cal{H}} = \bar{\cal{H}}\big(\bar{A}, \bar{a}_i, \bar{R}_{jk},\bar{\nabla}_{l}\big)$, where $\bar{A} = {A}(r)$ is the spherical seed solution of the gauge field. Then, 
taking Eqs.(\ref{2.2}) and (\ref{2.3})  into account,  we find that the Hamiltonian constraint  is 
satisfied identically to first-order $\omega$, if it is satisfied to its zeroth  order, even with the U(1) symmetry.

Because of the presence of $A$ and $\varphi$, the theory has two more  constraints in comparison to the one without the U(1) symmetry, obtained from 
the variation of the action with respect to $A$ and $\varphi$, given, respectively, by
\cite{WWa,ZWWS}
\bqn
\lb{equ12a}
&& R - 2\Lambda_g = 0,\\
\lb{equ12b}
&& {\cal{G}}^{ij} K_{ij} - \frac{1}{N}\left[\nabla_{i}\left(N a_j K^{ij}\right) - \nabla_{i}\left(N a^i K\right)\right]\nb\\
&& ~~~~~~ ~~ ~ + \frac{1-\lambda}{N}\left[\nabla^2\left(NK\right) - \nabla_{i}\left(N a^i K\right)\right] = 0, ~~
\eqn
where $  {\cal{G}}^{ij}  \equiv R^{ij} +(\Lambda_g - R/2)g^{ij}$. Note that the above constraints hold for both the projectable and non-projectable cases. 
Then, considering Eqs.(\ref{2.1a})-(\ref{2.3}) and the fact that $\bar{\nabla}_i\left(\bar{N}\bar{a}_j \delta{K}^{ij}\right) = 0$,   $\bar{g}_{ij}$ and $\bar{R}_{ij}$ are 
diagonal, one can see that the above equations are satisfied identically to the first-order of $\omega$, if they are satisfied to its zeroth-order.

With the gauge (\ref{gauge}), the momentum constraint takes the same form as that of Eq.(\ref{equ2}). As a result, to the first-order of $\omega$, it yields 
Eq.(\ref{equ4}).

The dynamical equations, on the other hand,  can be also cast in the form of Eq.(\ref{equ3}), if one replaces ${\cal{L}}_{K}$ by ${\cal{L}}_{K} + {\cal{L}}_{A}$, and 
now defines $F^{ij}$ as  $ F^{ij} \equiv  - \left[{\delta \left(N\sqrt{g}{\cal{L}}_{g}\right)}/{ \delta g_{ij}}\right]/({N\sqrt{g}})$,
where ${\cal{L}}_{g} \equiv {\cal{L}}_{V} + {\cal{L}}_{A} + {\cal{L}}_{\lambda,\varphi}$. Considering Eq.(\ref{gauge}), we find
$F^{ij} =  F^{ij}\big(a_{k}, R_{lm}, \nabla_{n}\big)$. Since ${\cal{L}}_{A}\left({A}, {R}\right) = \bar{\cal{L}}_{A}\left(\bar{A}, \bar{R}\right) + {\cal{O}}(\omega^2)$, and 
the rotation $\omega$ contributes to the dynamical  equations only through  the shift vector $N^i$ and extrinsic curvature tensor $K_{ij}$,
 one can easily show that to the first-order of $\omega$  only the first two terms in the right-hand of Eq.(\ref{equ3}) have
non-vanishing contributions, exactly   the same as those without the U(1) symmetry, and give rise to  two equations, given, respectively,  by Eqs.(\ref{equ5a}) 
and  (\ref{equ5b}).  Thus, in the present case we still have the same three field equations for $\omega(r, \theta)$ even with the enlarged symmetry (\ref{symmetry}). 
 Therefore, {\em if  $(N, f, h, A)$ is a vacuum solution of the HL theory with the local U(1) symmetry, either with the projectability condition or without it,
then, $(N, f, h, A, \omega)$ represents a corresponding slowly rotating vacuum solution of the same theory, where $\omega$ is given by Eq.(\ref{2.6})}. 
Since physics does not depend on the gauge choice, the above conclusion is also true in any gauge.  
 
{\em Conclusions}:---In this Letter, we have shown that, for any given spherical static vacuum solution of the HL theory, no matter it has the local U(1) symmetry 
or not, and whether it is with or without the projectability condition, there always exists a corresponding  solution, representing a slowly rotating vacuum space-time, 
which will reduce to the former when the rotation is switched off. 
 
It is remarkable to note that the rotation given by Eq.(\ref{2.6}) depends  on the models of the HL theory and their coupling constants only implicitly through 
the spherical seed solutions. All the  other effects are high orders of $\omega$. In this sense, the rotation is universal, and for all asymptotically flat seed solutions,
it takes precisely the form (\ref{2.7}) of the slowly rotating  Kerr solution of GR.

When specifying to the particular cases considered in \cite{GH},  we obtains the same results. On the other hand, our above conclusions are equally applicable to the 
IR limit of the non-projectable HL theory \cite{BPS}, since, as mentioned above, Eqs.(\ref{equ4}), (\ref{equ5a})  and  (\ref{equ5b}) do not contain  explicitly the 
coupling constants of the HL theory, and setting the ones that corresponds to the high-order derivative terms to zero 
will not affect the forms of these equations, although they do affect the spherical seed solutions $(N, f, h)$.    Recently  
spherical static vacuum spacetimes were studied  in  the IR limit of the non-projectable HL theory by using the equivalence between
it and the hypersurface-orthogonal Einstein-aether
theory \cite{Jacobson,BPS}, and a class of numerical solutions that represents black holes was found \cite{BJS}.   On the other hand,
Blas and Sibiryakov  also studied the same problem \cite{BSb},  and found that these black holes possess universal horizons.   In all of these studies,
  the Eddington-Finkelstein metric,
$
ds^2 = F(r) dv^2 - 2B(r)dv dr - r^2d\Omega^2,
$
 was used  \cite{BJS,BSb}, in which the four-velocity of the aether can be always parameterized as,
$u^{\alpha}\partial_{\alpha} = {\cal{A}}(r)\partial_{v} - {[1 - F(r) {\cal{A}}^2(r)]}/{[2  {\cal{A}}(r) B(r)]}\partial_{r}$, 
where $ {\cal{A}}$ is an arbitrary function, to be determined by the field
equations. 
In the spherical case,  since the aether is always  hypersurface-orthogonal \cite{Jacobson},
 one can introduce the time-like variable $t$, so that  
   $u_{\mu}$ takes the form $u_{\mu} = {t_{,\mu}}/{\left|g^{\alpha\beta}t_{,\alpha} t_{,\beta}\right|^{1/2}}$, from which 
we find that
$dv = {dt}/{t_{,v}} + {2 {\cal{A}}^2B}/{(1 +  {\cal{A}}^2F)} dr$.
The integrability  condition requires $t_{,vr} = 0$. Without loss of generality, we choose $t_{,v} = 1$, so  that
$v = t + \int^{r}{\left({2A^2B}/{(1 + A^2F)}\right)dr}$.
Inserting it into the Eddington-Finkelstein metric, we find that the corresponding ADM quantities are given by
Eq.(\ref{SRM}) with $\omega = 0$,
where 
$$
{N}^2 =  B^2 f=   \frac{(1+ {\cal{A}}^2F)^2}{4 {\cal{A}}^2},\;\;\;
{h} = \frac{1-  {\cal{A}}^4F^2}{4 {\cal{A}}^2B}. 
$$
Taking these black hole solutions as the seeds, from  the results presented 
above one can see that slowly rotation black  holes indeed exist in the IR limit of the non-projectable HL theory \cite{BPS}, 
in contrast to the claim presented in \cite{BS}. For detail, we refer readers to  \cite{Wang}.

It should be noted that the  equivalence between the hypersurface-orthogonal Einstein-aether theory and the IR limit of the non-projectable HL  theory
holds only in the level of action. In particular, the  Einstein-aether  theory still has the general diffeomorphisms as
that of GR, while the HL theory has only Diff($M, \; {\cal{F}}$). It is exactly because of the former that we are allowed to
make coordinate transformations of the kind $ t = t(v, r)$, which are forbidden by Eq.(\ref{0.2}). 

We also note that rotations (spins) of black holes are important not only because the observational fact that most of objects in our universe are rotating, 
as mentioned above, but also because  they might provide important information on the evolution histories  of  black holes and the formation of their jets \cite{Nara},
among other things. Indeed,
recently it was found  evidence that the relativistic jet of a black hole might be powered by its spin energy  \cite{NM}. Because of the universal 
form of the spin (\ref{2.7}), it might be difficult to distinguish the HL theory from others only by  measuring the spins of black holes 
(as far as slowly rotating black holes are concerned). However, a rotating 
source always drags space-time with it, and causes linearly polarized electromagnetic radiation to undergo polarization rotation - the gravitational Faraday effects
\cite{HD}. Lately, it was shown that this leads to a new relativistic effect that imprints orbital angular momentum on such light \cite{Obs}.
Since the resulted spectra depend not only on the spin of the black hole but also on the detailed structure (null geodesics) of the space-time,  they
might provide  important  information to distinguish different models of gravity, including the HL theory.

\section*{Acknowlodgements}
  
This work is supported in part by the DOE  Grant, DE-FG02-10ER41692. 


\end{document}